\pdfoutput=1
\documentclass[twocolumn,aps]{emulateapj09}

\usepackage{color}
\usepackage{graphicx, amsmath, amsmath, amssymb}
\usepackage[breaklinks, colorlinks, citecolor=blue]{hyperref}
\usepackage{listings} 
\newenvironment{mytabular}[1][1]{
  \renewcommand*{\arraystretch}{#1}
  \tabular
}{
  \endtabular
}

\def\aap{A\&A}

\def\apjl{ApJ}

\def\apjs{ApJS}

\newcommand{\adotoa}{\ensuremath{{\cal H}}}
\newcommand{\rhob}{\ensuremath{\bar{\rho}}}
\newcommand{\Pb}{\ensuremath{\bar{P}}}
\newcommand{\grad}{\ensuremath{\vec{\nabla}}}
\newcommand{\cv}{\ensuremath{c_{\text{vis}}}}

\shorttitle{Planck constraints on GDM}
\shortauthors{Thomas, Kopp, Skordis}

\begin{document}
\title{Constraining Dark Matter properties with Cosmic Microwave Background observations}
\author{Daniel B. Thomas}
\email{thomas.daniel@ucy.ac.cy}
\author{Michael Kopp}
\email{kopp.michael@ucy.ac.cy}
\author{Constantinos Skordis}
\email{skordis@ucy.ac.cy}
\affil{Department of Physics, University of Cyprus\\
1, Panepistimiou Street, 2109, Aglantzia, Cyprus}
\date{\today}

\begin{abstract}We examine how the properties of dark matter, parameterised by an equation of state parameter $w$
and two perturbative Generalised Dark Matter (GDM) parameters $c^2_s$ (the sound speed) and $c^2_\text{vis}$ (the viscosity), are
constrained by existing cosmological data, particularly the Planck 2015 data release.
We find that the GDM parameters are consistent with zero, and are strongly constrained, 
showing no evidence for extending the dark matter model beyond the Cold Dark Matter (CDM) paradigm. The dark matter equation of state
is constrained to be within  $-0.000896<w<0.00238$  at the $99.7\%$ confidence level, which is several times stronger than constraints found previously using
WMAP data. The parameters $c^2_s$ and $c^2_\text{vis}$ are constrained to be less than
$3.21\times10^{-6}$ and $6.06\times10^{-6}$ respectively at the $99.7\%$ confidence level. The inclusion of the
GDM parameters does significantly affect the error bars on several $\Lambda$CDM parameters,
notably the dimensionless dark matter density $\omega_g$ and the derived parameters $\sigma_8$ and $H_0$. This can be partially alleviated with the inclusion
of data constraining the expansion history of the universe.
 \end{abstract}


\section{Introduction}
The Lambda Cold Dark Matter ($\Lambda$CDM) cosmological model provides a good phenomenological fit to current cosmological data. It matches
the current expansion history of the universe including Supernovae \citep{PerlmutterAlderingGoldhaberEtal1999, RiessFilippenkoChallisEtal1998}
Baryon Acoustic Oscillation (BAO) \citep{AndersonAubourgBaileyEtal2014,RossSamushiaHowlettEtal2015} and local measurements
\citep{RiessMacriCasertanoEtal2011}, the matter power spectrum \citep{GilMarinNorenaVerdeEtal2015} and the Cosmic Microwave
Background (CMB) data from the Planck satellite \citep{PlanckCollaborationXIII2015}.

In this standard picture the CDM is a crucial component and is described as a non-interacting initially pressureless perfect fluid.
This perfect fluid has an equation of state parameter $w$ identically zero as well as zero sound speed and zero viscosity.
These CDM characteristics make it an extremely simple system to model physically, either perturbatively or through cosmological N-body simulations.

Astrophysical systems also provide important evidence for dark matter. Galaxy rotation curves \citep{PersicSalucciStel1996} were one of the earliest
suggestions of dark matter's existence \citep{RubinFord70,Zwicky1933}. Furthermore, gravitational-lensing measurements have been used to infer ``mass maps'' (see for example \cite{LeforFutamaseAkhlaghi2013}), that show the distribution
 of the CDM under the assumption that it traces the gravitational-lensing potential. These maps are consistent with what is expected from 
theory \citep{MasseyRhodesEllisEtal2007}. One of the most-studied gravitational-lensing systems is the ``bullet cluster'' 
\citep{CloweBradacGonzalezEtal2006}, a system consisting of two colliding galaxy clusters.
The ``bullet cluster'' system shows a clear displacement between the map of the gravitational-lensing potential and the map of the luminous matter, in a way that is
 naturally explained within the CDM paradigm. Since the discovery of the original system, further examples have been observed \citep{HarveyMasseyKitchingEtal2015}, 
providing more evidence in support of the CDM component of the standard cosmological model.

Despite the promising concordance of these results, no dark matter candidate has been found experimentally. Moreover, many physical models of dark matter are not as idealised as the pure CDM model. Thus, it seems
timely, with the precision cosmological data currently available, to examine constraints on departures from this idealised model.

Work has been carried out to constrain the properties of dark matter using astrophysical
systems, particularly galaxy rotation curves, see e.g. \cite{FaberVisser2006,SerraRomero2011,BarrancoBernalNunez2013,BoehmSchewtschenkoWilkinsonEtAl2014}.
Interestingly, it has been also pointed out that halo properties deviate from expectations of $\Lambda$CDM \citep{Moore1994,JeeHoekstraMahdaviEtal2014,BoylanKolchinBullockKaplinghat2011,
PapastergisGiovanelliHaynesEtal2015}. As we are only interested in the properties of dark matter on linear scales in this paper, we will not discuss halos further.

Cosmological constraints on properties of dark matter beyond CDM have been investigated within the context of specific models.
In \cite{Yang2015}, Planck data was used to constrain the properties of dark matter, focusing on the dark matter mass, decay rate and the thermally averaged cross section for annihilation.
CMB data from Planck as well as large scale structure observations have been used to place constraints on
interactions of dark matter with other Standard Model (SM) particles, for instance, a possible dark matter-photon~\citep{WilkinsonLesgourguesBoehm2014}
 or a dark matter-neutrino~\citep{WilkinsonBoehmLesgourgues2014} interaction. The elastic scattering of dark matter with other SM particles 
also leaves an imprint on spectral distortions on the CMB, as investigated lately in \cite{Ali-HaimoudChlubaKamionkowski2015}, where constraints were put on the interaction cross-section.
Dark matter may also interact with a dark radiation component \citep{Buen-AbadMarques-TavaresSchmaltz2015,LesgourguesMarques-TavaresSchmaltz2015}
which may further have its own dark recombination and dark atomic structure \citep{Cyr-RacineSigurdson2012}.
Other well motivated models are axions \citep{HlozekGrinMarshEtal2015},
collisionless warm dark matter \citep{Armendariz-PiconNeelakanta2014,PiattellaCasariniFabrisEtal2015} and collisionless massive neutrinos \citep{ShojiKomatsu2003}.

In this paper we are interested in constraining the properties of dark matter by modelling the dominant component of structure formation beyond a simple pressureless perfect fluid. 
This could either be due to a more complicated dark matter model, or just a more precise modelling of CDM, as in the case of the
 Effective Field Theory of Large Scale Structure (EFTofLSS) \citep{BaumannNicolisSenatoreEtal2012}. 
For convenience throughout this paper, we consider ``CDM''
 to be the modelling of the dark matter component as a pressureless perfect fluid. 
Thus, both more complicated models or a more precise modelling of the evolution of CDM are considered to be ``beyond CDM'', in the sense that they 
are a change to how the dark matter is usually modelled. 

 We parameterise the departures from CDM according to the Generalised Dark Matter (GDM) model \citep{Hu1998a}. 
As we discuss in more detail in a companion paper  \citep{KoppSkordisThomas2015}, the GDM parameterisation naturally arises in more realistic models, 
for instance, the  EFTofLSS \citep{BaumannNicolisSenatoreEtal2012}, non-equilibrium
thermodynamics \citep{LandauLifshitz1987}, the effective theory of fluids \citep{Ballesteros2015}, tightly coupled fluids and scalar fields.
In \cite{KoppSkordisThomas2015} we also study more closely the physical effects and the interpretation of the GDM parameters.

Several authors have used the GDM model to put cosmological constraints on dark matter properties
  \citep{Muller2005,CalabreseMigliaccioMelchiorriEtal2009,LiXu2014,WeiChenLiu2013,KumarXu2012,Xu2014,XuChang2013}, as we also do in this paper.
We perform a more detailed comparison to these works in section \ref{sec_comparison}.
Apart from describing dark matter, the GDM parameterisation or a subset of it, has been used in some form in the literature for several different purposes 
including neutrinos \citep{TrottaMelichiorri2005}, dark energy \citep{WellerLewis2003,BeanDore2004} 
and unified dark matter/dark energy models \citep{KumarXu2012,YangXu2013}. 

This paper is organised as follows. Section \ref{sec_theory} contains a brief introduction to the GDM model. Section \ref{sec_method}
details the method and data that were used in our analysis, and our results are presented in section \ref{sec_results}. We conclude
in section \ref{sec_conclusion}.

\section{A short overview of the GDM model}
\label{sec_theory}
In this section we present the basic ingredients of the GDM model~\citep{Hu1998a}. We work in the synchronous gauge as this is the gauge most commonly used in numerical Boltzmann codes. We are interested only in scalar perturbations, so the metric takes the form
\begin{equation}
ds^2 = a^2 \left\{ - d\eta^2 + \left[(1 + \frac{1}{3} h) \gamma_{ij} + D_{ij} \nu \right] dx^i dx^j \right\} \,
\end{equation}
where $a$ is the scale factor in conformal time $\eta$,  $\gamma_{ij}$ is a flat spatial metric with covariant derivative $\grad$,
$h$ and $\nu$ are the two scalar metric perturbations in this gauge and $D_{ij} = \grad_i \grad_j - \frac{1}{3} \grad^2 \gamma_{ij}$ is a traceless spatial operator.
The GDM has background density $\rhob_g$ and isotropic pressure $\Pb_g$, related by an equation of state $w$, such that
  
\begin{align}
\dot{\rhob}_g & = - 3 \adotoa (1+w) \rhob_g \\
w &= \frac{\Pb_g}{\rhob_g}\,.
\end{align}
  
Unlike CDM, the GDM is allowed to have a pressure perturbation $\Pi_g$ and shear perturbation $\Sigma_g$ \citep{Hu1998a}, in addition
to the usual density ($\delta_g$) and velocity ($\theta_g$) perturbations. The
perturbations obey the Euler and continuity equations, as well as two postulated closure equations for the pressure perturbation and the shear \citep{Hu1998a},
see also \cite{KoppSkordisThomas2015} for an extended discussion and an explanation of our notation. The equations that the GDM obeys in a spatially flat Universe are as follows
\begin{align}
 \dot{\delta}_g  &= 3  \adotoa \left( w \delta_g  - \Pi_g\right) -  (1 + w) \left( \frac{1}{2} \dot{h}  - \grad^2\theta_g \right) \\
 \dot{\theta}_g  &=  (3 c_a^2 - 1)     \adotoa   \theta_g +  \frac{\Pi_g}{1+w} +  \frac{2}{3}  \grad^2  \Sigma_g \\
\dot{\Sigma}_g &=   - 3 \adotoa \Sigma_g+ \frac{4}{(1+w)} c_{\rm vis}^2 (\theta_g - \frac{1}{2}\dot{\nu} ) \\
 \Pi_g  &=   c_s^2 \delta_g +  3 \adotoa (1+w)( c_s^2 - c_a^2 )  \theta_g \,,
\end{align}
where we have introduced two sound speeds and a viscosity (the equivalent equations in a spatially curved Universe and in a general gauge can be found in \cite{KoppSkordisThomas2015}). The sound speed $c^2_s$ and viscosity $c^2_\text{vis}$ are parameters of the GDM model, and the adiabatic
sound speed  $c_a^2$ is defined as
\begin{equation}
 c_a^2 = \frac{\dot{\Pb}_g}{\dot{\rhob}_g} = w - \frac{\dot{w}}{3 \adotoa(1+w)}\,.
 \end{equation}
In the present work we consider only constant equation of state $w$ so that $c_a^2 = w$.

In general, the sound speed $c^2_s$ causes oscillations in the density perturbation $\delta_g$ below the Jeans length, although if $c^2_s$ and $c^2_\text{vis}$ become comparable in size then there are no sound waves. For convenience, we refer to $c^2_s$ as the sound speed, although this is only true for $c^2_s \gg c^2_\text{vis}$ \citep{KoppSkordisThomas2015}. The viscosity $c^2_\text{vis}$
damps the density perturbations. For details of the model and investigations into which physical models map to GDM, see \cite{Hu1998a,KoppSkordisThomas2015}.

To summarize, replacing CDM with GDM amounts to introducing three parametric functions to the model: the GDM equation of state $w(\eta)$, the sound speed $c_s^2(\eta,\vec{x})$ and 
the viscosity $\cv^2(\eta,\vec{x})$. In this work we shall assume that all three take constant values, i.e.  with no time or space dependence.
 Whilst this is not the case for many of the physical models that GDM relates to, these constant parameters
can be considered as a null test as to whether there is any evidence for departures from CDM. 
\begin{table*}
 \caption{Constraints on the GDM parameters for the two types of models and different combinations of experiments, for the $95.5\%$ and $99.7\%$ credible regions.}
\centering
\begin{mytabular}[1.8]{|c||cc||cc|cc|cc||} 
 \hline 
 \hline 
 &\multicolumn{2}{|c||}{ $\Lambda$-wDM}& \multicolumn{6}{|c||}{$\Lambda$-GDM}  \\
 \cline{2-9}
 &\multicolumn{2}{|c||}{  $10^2 w$}  & \multicolumn{2}{|c|}{$10^2w$}  & \multicolumn{2}{|c|}{$10^6c_s^2$, upper bounds}  & \multicolumn{2}{|c||}{$10^6\cv^2$, upper bounds}   \\
Likelihoods &   $95.5\%$ & $99.7\%$ &  $95.5\%$ & $99.7\%$  &  $95.5\%$ & $99.7\%$ &  $95.5\%$ & $99.7\%$  \\
\hline 
PPS & $0.007^{+0.463}_{-0.466}$ & $0.007^{+0.676}_{-0.673}$ & $-0.040^{+0.473}_{-0.468}$  & $-0.040^{+0.700}_{-0.701}$   & $ 3.31$ & $ 6.31$ &  $ 5.70$ & $ 11.3$   \\
PPS + Lens & $0.087^{+0.439}_{-0.448} $ & $0.087^{+0.662}_{-0.648}$ &  $0.066^{+0.434}_{-0.427} $ & $0.066^{+0.654}_{-0.642}$ & $ 1.92$ & $ 3.44$ & $ 3.27$ & $ 5.99$  \\
PPS + Lens + HST & $0.256^{+0.217}_{-0.217} $ & $0.256^{+0.322}_{-0.323}$ & $0.259^{+0.216}_{-0.218} $ & $0.259^{+0.321}_{-0.326}$ &  $ 1.87$ & $ 3.38$ &  $ 3.11$ & $ 5.56$ \\
PPS + Lens + BAO & $0.063^{+0.108}_{-0.112} $ & $0.063^{+0.163}_{-0.164}$ & $0.074^{+0.111}_{-0.110} $ & $0.074^{+0.164}_{-0.163}$ &  $ 1.91$ & $ 3.21$ &  $ 3.30$ & $ 6.06$  \\
\hline 
 \hline 
 \end{mytabular} 
\label{table_w_speeds}
\end{table*}

\section{Method and Data}
\label{sec_method}
In order to perform our analysis, we modified the Cosmic Linear Anisotropy Solving System (CLASS) code \citep{Lesgourgues2011}. CLASS numerically solves the Boltzmann equation for each 
relevant component coupled to the Einstein equations and calculates the CMB and matter power spectra given a set of model parameters.
 The CLASS code already includes an additional dark energy fluid component 
with an equation of state and sound speed \citep{LesgourguesTram2011} which we further modified to include the viscosity $\cv^2$ and to allow this fluid to work as a 
replacement for dark matter rather than for dark energy. We also independently modified a different Boltzmann code (DASh) \citep{KaplighatKnoxSkordis2002} to include the 
full GDM parameterisation. We performed a full comparison between the codes, including the background evolution, perturbation evolution, the $C_l$s, matter power 
spectrum and lensing potential. The numerical difference of the two codes in the case of the GDM model is similar to the corresponding difference in the case of $\Lambda$CDM, within 
$\sim 0.1\%$.
 This level of agreement holds for all quantities in both the synchronous gauge and the conformal Newtonian gauges.~\footnote{The actual difference 
between the codes in the case of $\Lambda$CDM ranges from around $10^{-4}$  on small scales to around $10^{-3}$ on large scales.
However, as DASh is an older code and not as optimized as CLASS we believe CLASS to be more accurate.}

We investigated the constraints on the GDM parameters using a Markov Chain Monte Carlo (MCMC) approach, carried out using the publicly available MontePython 
code \citep{AudrenLesgourguesBenabedetal2013}, which implements the Metropolis-Hastings algorithm.
The MontePython code calls the CLASS code through a Python wrapper. 
Our covariance matrix for the final runs was generated using the standard methodology: first some initial runs were carried out with a simple estimated diagonal covariance matrix, and these were used to generate a covariance matrix that was used for the next run. After several iterations of this we had a covariance matrix that gave an appropriate acceptance rate for the steps in the chains and was suitable for the final runs.

We considered three types of models: the standard  $\Lambda$CDM model (used for comparison), the  $\Lambda$-wDM model where CDM is replaced with GDM but the speed of sound and viscosity
are set to zero, and the $\Lambda$-GDM model where all three GDM parameters are included. Note that in all three models, the dark energy component is always modelled as a cosmological constant
$\Lambda$ and only the dark matter component is modified. 

In all models we varied the standard $\Lambda$CDM parameters with the MCMC algorithm:
 the dark matter dimensionless density $\omega_g$,\footnote{For  $\Lambda$CDM, the parameter $\omega_g$ is equal to $\omega_c$ the CDM dimensionless density, as in that case
 $w=c^2_s=c^2_\text{vis}=0$.} 
the baryon dimensionless density $\omega_b$, $100\times\theta_s$ where $\theta_s$ is the ratio of the sound horizon to the angular diameter distance at decoupling, the optical depth $\tau$,
$\ln(10^{10}A_{s })$ where $A_s$ is the amplitude of scalar perturbations and the spectral index of scalar perturbations $n_s$ (6 parameters total). 
For the $\Lambda$-wDM model we varied in addition to the $\Lambda$-CDM  parameters the GDM equation of state $w$ (7 parameters total) and for
the  $\Lambda$-GDM model we further varied the sound speed $c_s^2$ and the viscosity $\cv^2$ (9 parameters total). 
In addition to the main MCMC parameters we considered three more derived parameters: the Hubble constant $H_0$ (in units of $\mathrm{km/s/Mpc}$),
the cosmological constant relative density $\Omega_\Lambda$ and the RMS matter density fluctuation $\sigma_8$. 
We used flat priors on all of the parameters and restricted $c^2_s$ and $\cv^2$ to be non-negative as dictated by theoretical models \citep{KoppSkordisThomas2015}, and $\tau$ to be greater than $0.01$. 
The remaining fiducial cosmology was set as follows. The spatial curvature was set to zero with a cosmological constant $\Omega_{\Lambda}$ making up the remainder of the matter content
and the primordial helium fraction was set to $Y_{\text{He}}=0.2477$\footnote{
The value for $Y_{\text{He}}$ is the (rounded) value quoted by Planck 2013 \citep{PlanckCollaborationXVI2013}. 
 It was verified by a preliminary Fisher-matrix analysis that including  $Y_{\text{He}}$
as a parameter would lead to minimal changes to the constraints on the GDM parameters.}. 
We used two massless and one massive neutrino with mass $0.06$ eV keeping the effective number of neutrinos to $N_\mathrm{eff} = 3.046$. \footnote{
In CLASS, this required us to set the effective number of massless neutrinos parameter to $N_{\text{ur}}=2.0328$ and 
the neutrino temperature parameter to $T_\text{ncdm}=0.71611$. Note that this is slightly larger than the 
instantaneous decoupling value ratio to the photon temperature of $(4/11)^{1/3}$, see the CLASS explanatory parameter file or \cite{ManganoMielePastorEtAl2005} for details.}

The main dataset that we used was the Planck 2015 data release \citep{PlanckCollaborationXI2015} of the CMB anisotropies. 
We used the low-$l$ likelihood and the full TT, EE and TE high-$l$ likelihood with the complete ``not-lite'' set of nuisance
parameters. The low-$l$ likelihood consists of the TT, EE, TE and BB spectra up to $l=29$, whereas the high-$l$ spectra are from $l=30$ upwards. See the Planck papers and wiki\footnote{
http://wiki.cosmos.esa.int/planckpla2015/index.php/
}
for full details of these likelihoods. As we always used the high-$l$ and low-$l$ likelihoods together, this combination will be referred to simply as Planck Power Spectrum (PPS). 
We included Gaussian priors on the nuisance parameters (also varied as MCMC parameters) as recommended by the Planck collaboration and implemented in the MontePython code. 
We ran further chains that included the Planck lensing potential likelihood (hereby referred to as ``Lens'') in addition to the low-$l$ and high-$l$ likelihoods, 
and found that this made a significant difference to the constraints on $c^2_s$ and $\cv^2$. 
Finally, we also investigated the effect on the constraints of other cosmological datasets that constrain the expansion history of the universe. 
The two datasets used for this were the HST key project \citep{RiessMacriCasertanoEtal2011} 
and Baryon Acoustic Oscillation (BAO) data \citep{AndersonAubourgBaileyEtal2014,BeutlerBlakeCollessEtAl2011}. 
The HST likelihood was implemented by a Gaussian prior on $H_0$ around  $H_0 = 73.8 \pm 2.4$, whereas the BAO data constrains the distance combination
\begin{equation}
D_V(z)=\left[cz(1+z)^2D^2_A(z)H^{-1}(z) \right]^{\frac{1}{3}}\text{,}
\end{equation}
where $D_A(z)$ is the angular diameter distance to redshift $z$, $H(z)$ is the Hubble parameter and $c$ is the speed of light. 
\begin{figure}
  \centering
 \includegraphics[width=3.4in]{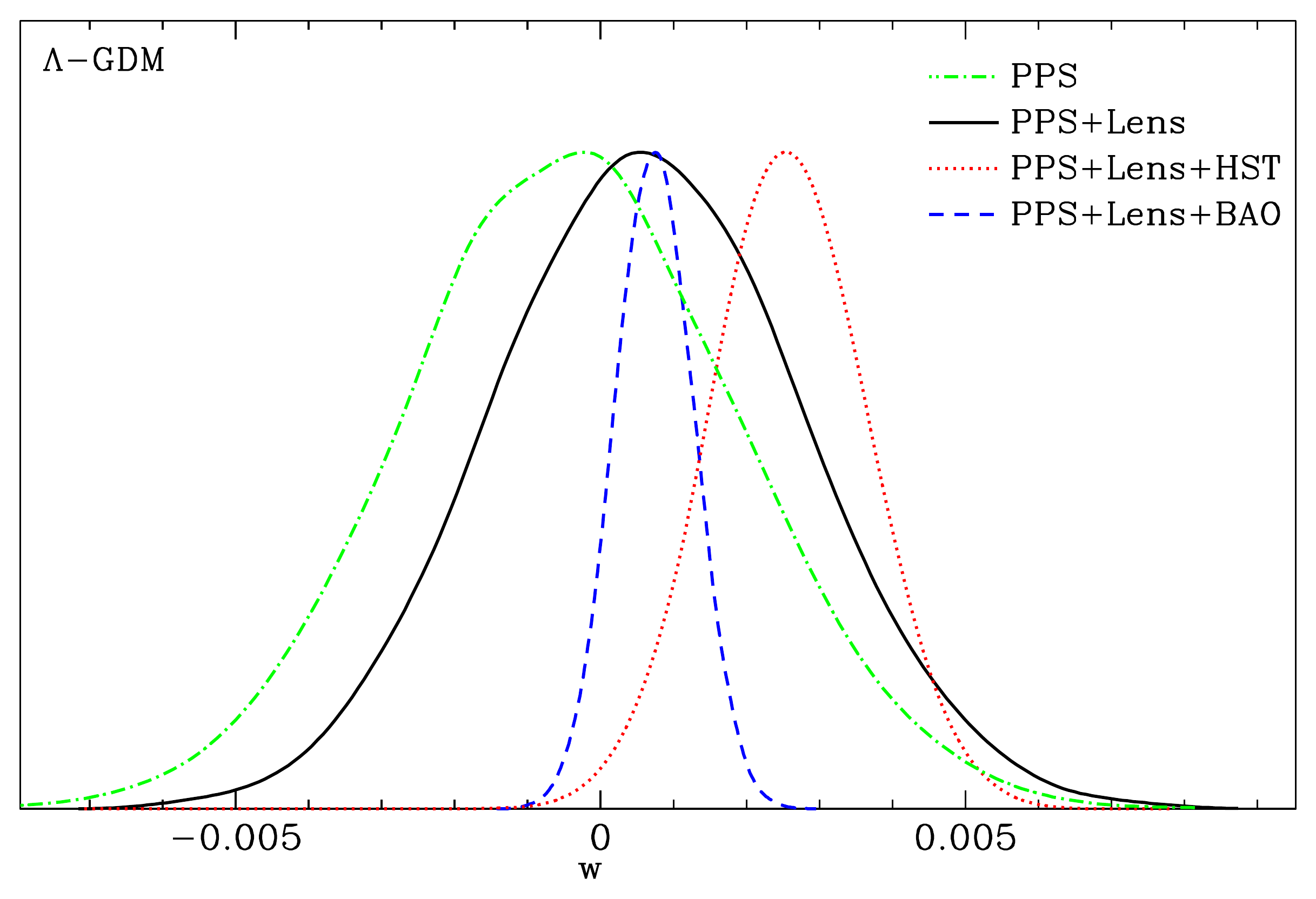}
\caption{The 1d posteriors for the $w$ parameter within the $\Lambda$-GDM model. 
The green (dot-dashed) curve is for the Planck Power Spectrum (PPS) dataset, the black (solid) curve is for the combination of PPS and Planck lensing (Lens),
 the red (dotted) curve is for the PPS and Lens dataset combination with the addition of HST prior and the blue (dashed) curve is for the PPS, Lens and BAO dataset combination.
}
\label{fig_1d_w}
\end{figure}

For the final results we generated a set of 8 chains for each dataset combination within each of the three models considered. Convergence of the chains was tested using 
the Gelman-Rubin $1-R$ \citep{GelmanRubin1992} test, which compares the variance within each chain to the variance between the chains in order to assess whether the individual chains have converged to the same posterior distribution. Convergence was determined by requiring  $1-R$ to be smaller than $0.01$ for all parameters. 

Note that GDM is defined for linear perturbations only, thus we have made no use of halofit to model non-linearities. In $\Lambda$CDM, halofit makes a small difference to the lensing potential and thus also to the lensed $C_l$s\footnote{We thank Steffen Hogstatz for bringing this to our attention.}. 
For parameter values around the $99.7\%$ confidence level (CL) of our constraints, the GDM effects are approximately four times larger than the effect 
of halofit for $\Lambda$CDM, but in the opposite direction. Thus, we expect that the constraints would not undergo a significant change 
if a full non-linear analysis was performed.
\begin{figure}
\centering
 \includegraphics[width=3.4in]{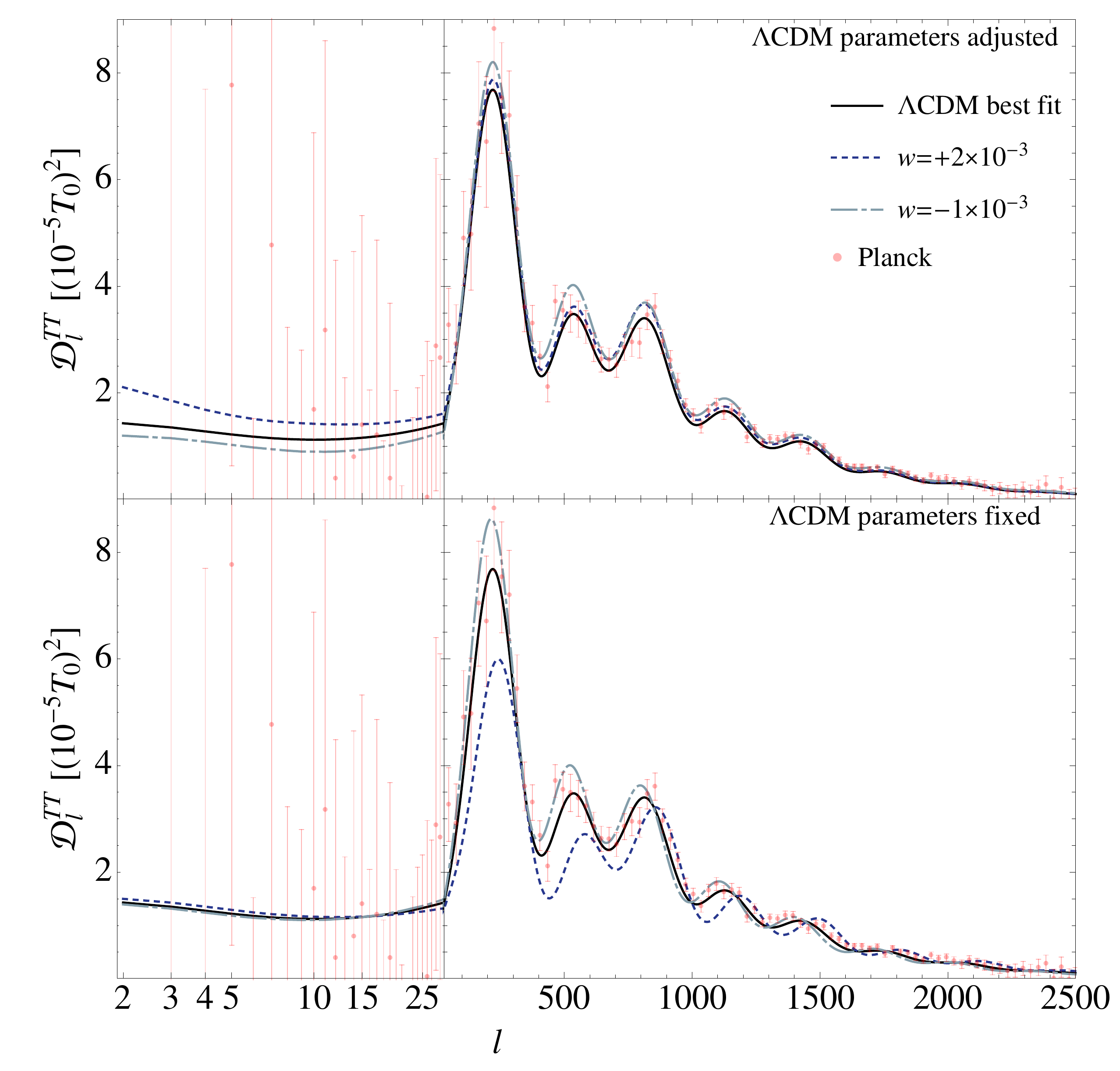}
 \caption{The $\Lambda$CDM best-fit $C_l$s along with $C_l$s plotted for two values of $w$, close to our $99.7\%$ CL values. In the lower panel, the standard cosmological parameters are held at the $\Lambda$CDM best-fit values when $w$ is varied, 
and in the upper panel they are varied in order to mask the effect of $w$. All curves as well as the Planck data points and their error bars
have been transformed by $C_\ell \rightarrow C_\ell + 10(C_\ell^{\Lambda \rm{CDM}} - C_\ell)$, effectively enlarging the residuals by 10
and shifting the mean values accordingly, to make them more visible.
 Comparison of the upper and lower plots show how the degeneracies that $w$ has with the standard cosmological parameters act to limit the constraints on $w$.
 }
\label{fig_w_clchanges}
\end{figure}

\section{Results}
\label{sec_results}
\subsection{Constraints on the GDM parameters}
Our main results are displayed in table \ref{table_w_speeds}, where we tabulate the $95.5\%$ and $99.7\%$ credible regions for $w$, $c^2_s$ and $\cv^2$
for the $\Lambda$-wDM and  $\Lambda$-GDM models for each of the different choices of datasets: PPS, PPS+Lens, PPS+Lens+HST and PPS+Lens+BAO. 
The constraints on the common parameters to $\Lambda$CDM are displayed in table \ref{table_common_parameters}. 

The first result to note from table \ref{table_w_speeds} is that all three GDM parameters are consistent with zero in all cases. The constraints on the GDM parameters are strong regardless of the dataset combination used. The strongest constraints come from the PPS+Lens+BAO combination where $-0.00109<w<0.00225$ at the $99.7\%$ CL in the case of  $\Lambda$-wDM 
and  $-0.000896<w<0.00238$ at $99.7\%$ CL in the case of  $\Lambda$-GDM. For the latter model the parameters $c^2_s$ and $\cv^2$ are constrained to be less than
$3.21\times10^{-6}$ and $6.06\times10^{-6}$ respectively at the $99.7\%$ CL. The 1-D posteriors for the $w$ parameter in the  $\Lambda$-GDM model are plotted
in Fig.~\ref{fig_1d_w} for all four combinations of datasets.  We now examine more closely how these constraints vary between the different choices of experiment.
\begin{figure}
\centering
 \includegraphics[width=3.4in]{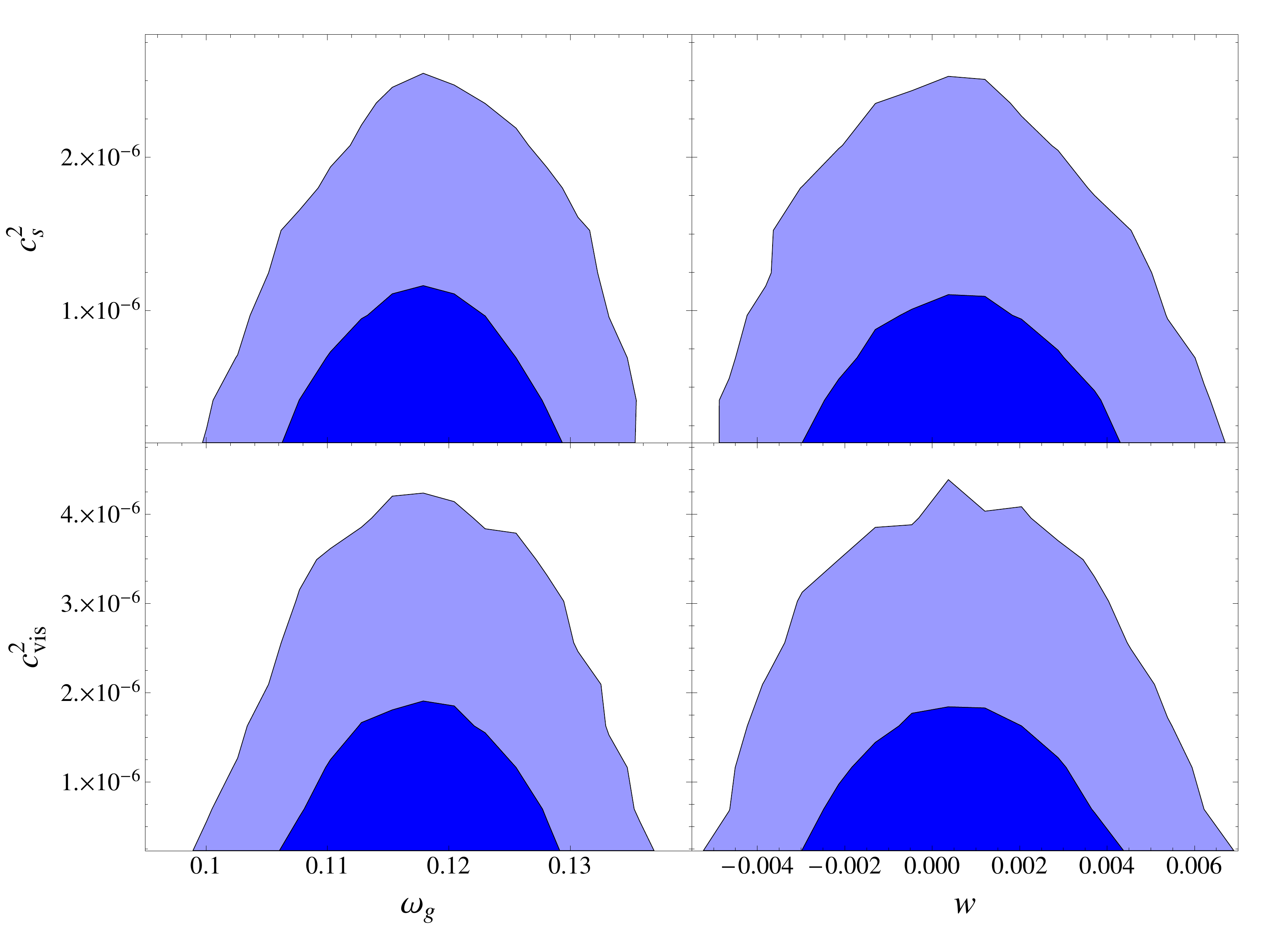}
 \caption{ The $68.3\%$ and $95.5\%$ credible regions for combinations of parameters within the  $\Lambda$-GDM model for the PSS+Lens dataset combination. The displayed combinations are:
 contours for the $\omega_g$-$c^2_s$ combination (upper left); contours for the $\omega_g$-$\cv^2$ combination (lower left);
 contours for the $w$-$c^2_s$ combination (upper right); contours for the $w$-$\cv^2$ combination (lower right).
The lack of correlation shown in these plots indicates a clear split between the perturbative GDM parameters and background GDM parameters. 
 }
\label{fig_2d_w_cs2}
\end{figure}

Consider first the constraints from the Planck Power Spectra. As discussed in \cite{Hu1998a} and \cite{KoppSkordisThomas2015}, the main effect of increasing (decreasing) $w$ 
is to shift the radiation-matter equality to earlier (later) times which in turn decreases (increases) the acoustic driving effect caused by the time-variation of the potential wells. 
The result is a decrease (increase) of the anisotropies around the first and second peak. As further explained in \cite{KoppSkordisThomas2015} potential decay after recombination
is related only to a time varying \emph{total} equation of state (that is, the combined equation of state from all species), hence, changing $w$ has only 
a very mild effect on the early Integrated Sachs-Wolfe (ISW) term and the lensing potential. 
This means that the ISW effects on the CMB TT and TE power spectra and the lensing effects on all the CMB power spectra are of similar strength to $\Lambda$CDM 
and do not drive the constraints on $w$. Finally, changing $w$ results in a change to $\theta_s$ that induces a lateral shift in the location of the CMB peaks, although
this is overshadowed by the changes of the peak heights. 

In figure \ref{fig_w_clchanges}, we plot the temperature $C_l$s for the best-fit $\Lambda$CDM parameters, plus two curves with $w$ values around our $99.7\%$ CL limits. In the bottom panel, the standard 6 $\Lambda$CDM parameters are held fixed when $w$ is varied. However, in the upper panel all of the $\Lambda$CDM parameters were varied in addition to $w$, guided by the results from our chains. Comparing the top and bottom plots, the importance of the degeneracies of $w$ with the standard cosmological parameters can be seen: these degeneracies act to mask the effects of $w$. Thus, if the standard cosmological parameters were known, then the constraints on $w$ would be significantly tighter. 
Note that in these plots the residuals of the data points and the $w$ curves with respect to the $\Lambda$CDM best-fit 
curve have been multiplied by 10 in order to make the differences clearer, by using the transformation $C_\ell \rightarrow C_\ell + 10(C_\ell^{\Lambda \rm{CDM}} - C_\ell)$.

The addition of CMB lensing to the temperature and polarisation $C_l$s has little effect on the posterior distribution for $w$, other than a slight shift of the mean to positive values without changing the width of the distribution. This is not surprising as the lensing is generally less constraining than the other $C_l$s from Planck due to the larger noise. Furthermore, although $w$ does have some effects on the lensing potential, such as causing the gravitational potentials to freeze out at different values (see \cite{KoppSkordisThomas2015}), these effects can be partially compensated as for the temperature $C_l$s in figure \ref{fig_w_clchanges}.

The addition of the data related to the expansion history does have a strong affect on $w$, as expected. The $95.5\%$ and $99.7\%$ CL 
constraints on $w$ are tighter by up to a factor of two when the HST prior is included. In addition, the mean value is also increased to a value just outside 
of the $68.3\%$ credible region, although this is not strongly significant. Note that in this case, $w=0$ is excluded at the $95.5\%$ CL. There is a known tension between
the $H_0$ value from Planck and the HST prior, so that one cannot yet safely conclude that this is a signature of new physics. 
We have repeated this run using the alternative value $H_0=70.6\pm3.3$ from the analysis in \cite{Efstathiou2015}, with the result that the posterior shifts to the left and widens slightly compared to the standard HST value. In this case $w$ is consistent with zero.

The change to the mean value does not appear when BAO data is used instead of the HST prior. 
Furthermore, with the addition of the BAO data to the Planck combination of PPS+Lens, the constraints on $w$ are improved by a factor of four. These are the tightest 
constraints on $w$ presented in this paper. The 1-D posteriors for $w$ for the $\Lambda$-GDM model can be seen in Fig.~\ref{fig_1d_w}, where the green (dot-dashed) 
 shows the posterior for PPS only, the black (solid) curve for the PPS+Lens combination 
and the red  (dotted) and  blue (dashed) curves show the posteriors for the addition of the HST and BAO data 
respectively. The changes to the constraints, and the shift of the peak when the HST data is included, are all clearly visible.

\begin{figure}
  \centering
  \hspace{-0.2in}
\includegraphics[width=3.4in]{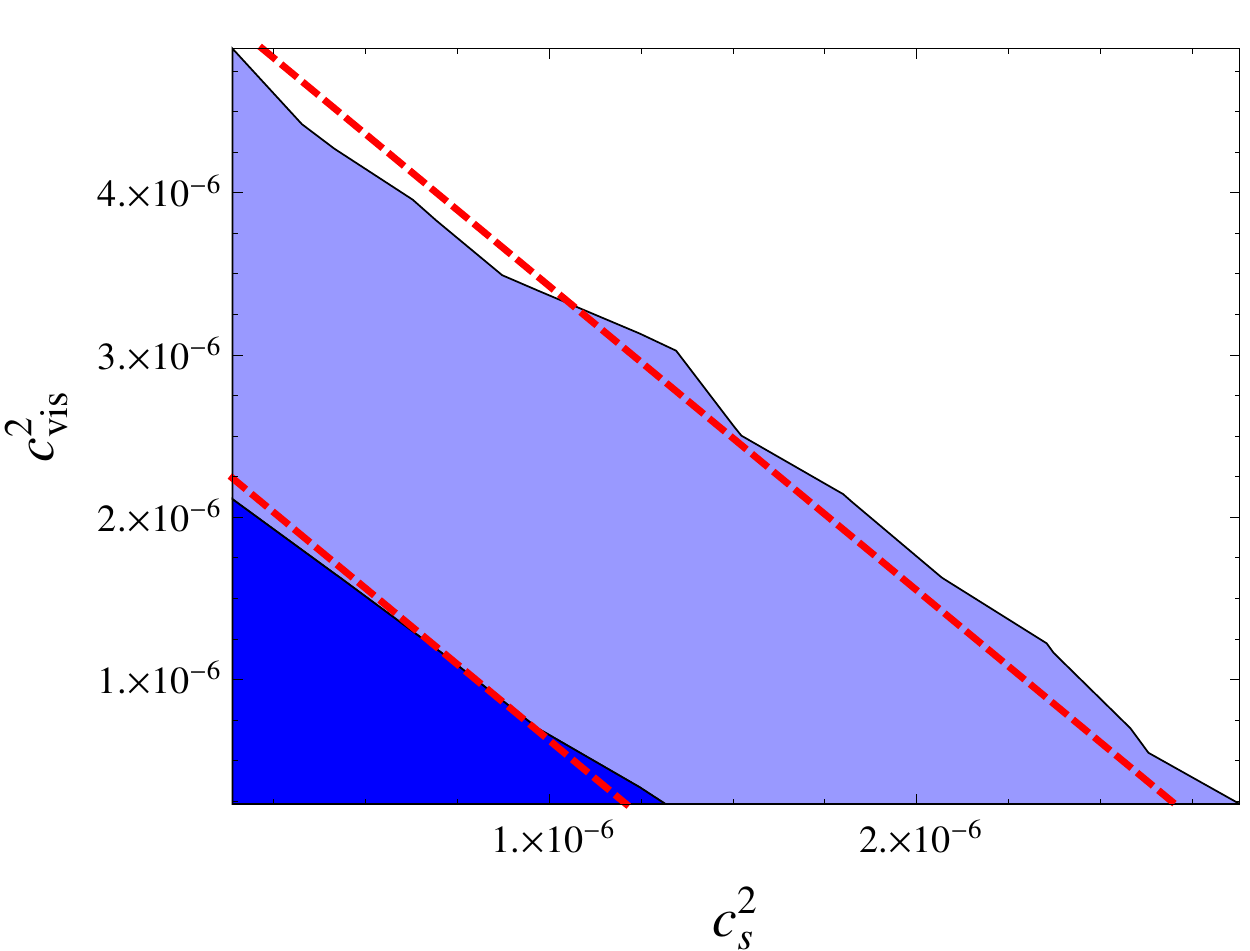}
 \caption{The $68.3\%$ and $95.5\%$ credible regions for $c^2_\text{vis}$ and $c^2_s$, showing the strong negative correlation between these parameters. These contours are for the $\Lambda$-GDM run for Planck+Planck lensing. The red (dashed) lines denote
 lines of constant $c^2_s+\frac{8}{15}c^2_\text{vis}$, showing that the degeneracy is well fit by the $k_\text{d}$ parameter as explained in the text.}
\label{fig_2d_cv2_cs2}
\end{figure}

We now turn to the $\Lambda$-GDM model and in particular on the constraints to the GDM  perturbative parameters $c_s^2$ and $\cv^2$. Firstly, the constraints on $w$ are not significantly
affected by the inclusion of the other GDM parameters,  $c_s^2$ and $\cv^2$, as can be seen by the similarity of the constraints on $w$
 between the $\Lambda$-wDM and $\Lambda$-GDM models for each dataset combination. This is to be expected as our discussion above shows that $w$ affects the CMB differently 
than the two perturbative GDM parameters. In the right panels of Fig.~\ref{fig_2d_w_cs2} we show the 2d-contours in the $w$-$c_s^2$ (upper right) and $w$-$\cv^2$ (lower right) planes respectively for the 
PPS+Lens dataset combination. The other dataset combinations give similarly looking contours. These plots show that $w$ is not strongly correlated with either of the other GDM parameters. This lack of correlation
indicates a clear split between the perturbative GDM parameters and background GDM parameters. 

\begin{figure}
\centering
 \includegraphics[width=3.4in]{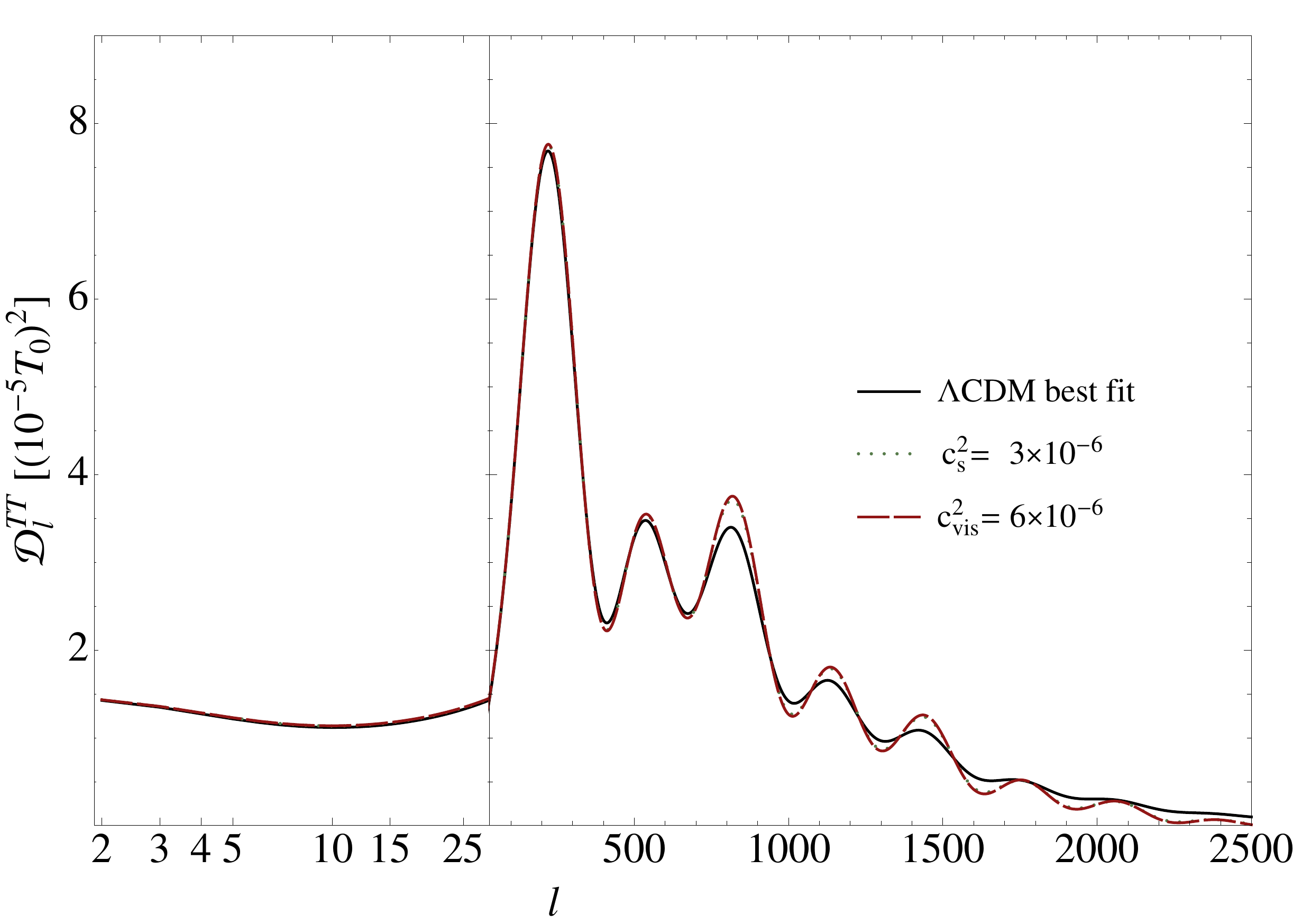}
 \caption{The TT spectrum plotted for the $\Lambda$CDM best-fit values and for values of $c^2_s$ and $c^2_\text{vis}$ that are close to our $99.7\%$ CL. The residuals of the $c^2_s$,$c^2_\text{vis}\neq0$ curves with respect to the best-fit $C_l$s have been multiplied by a factor of 50 to make them more visible as in Fig.\ref{fig_w_clchanges}. The key difference between the GDM and $\Lambda$CDM curves is the reduced smoothing of the peaks in the GDM model, caused by the reduced lensing potential.
 }
\label{fig_cscv_TTchanges}
\end{figure}

Moving on to the perturbative GDM parameters themselves, the main effect of non-zero $c_s^2$ and $\cv^2$, as found in \cite{KoppSkordisThomas2015},
is to cause the gravitational potential to decay.  Potential decay results in two main effects on the CMB power spectrum which drive the constraints on $c_s^2$ and $\cv^2$. The first is a
continuous ISW effect after recombination until the present time, which becomes stronger with increasing values of the perturbative GDM parameters.
The second effect is on the lensing potential, also examined in \cite{KoppSkordisThomas2015}. Since the lensing potential is directly sourced by the gravitational potential,
potential decay leads to a smaller CMB lensing signal on the CMB power spectra.  As the CMB lensing decreases the peak heights and increases the peak troughs (without changing their location), the
reduced lensing potential in $\Lambda$-GDM results in higher peak heights and lower peak troughs compared to either $\Lambda$-wDM or $\Lambda$CDM models.
\begin{figure}
\centering
 \includegraphics[width=3.4in]{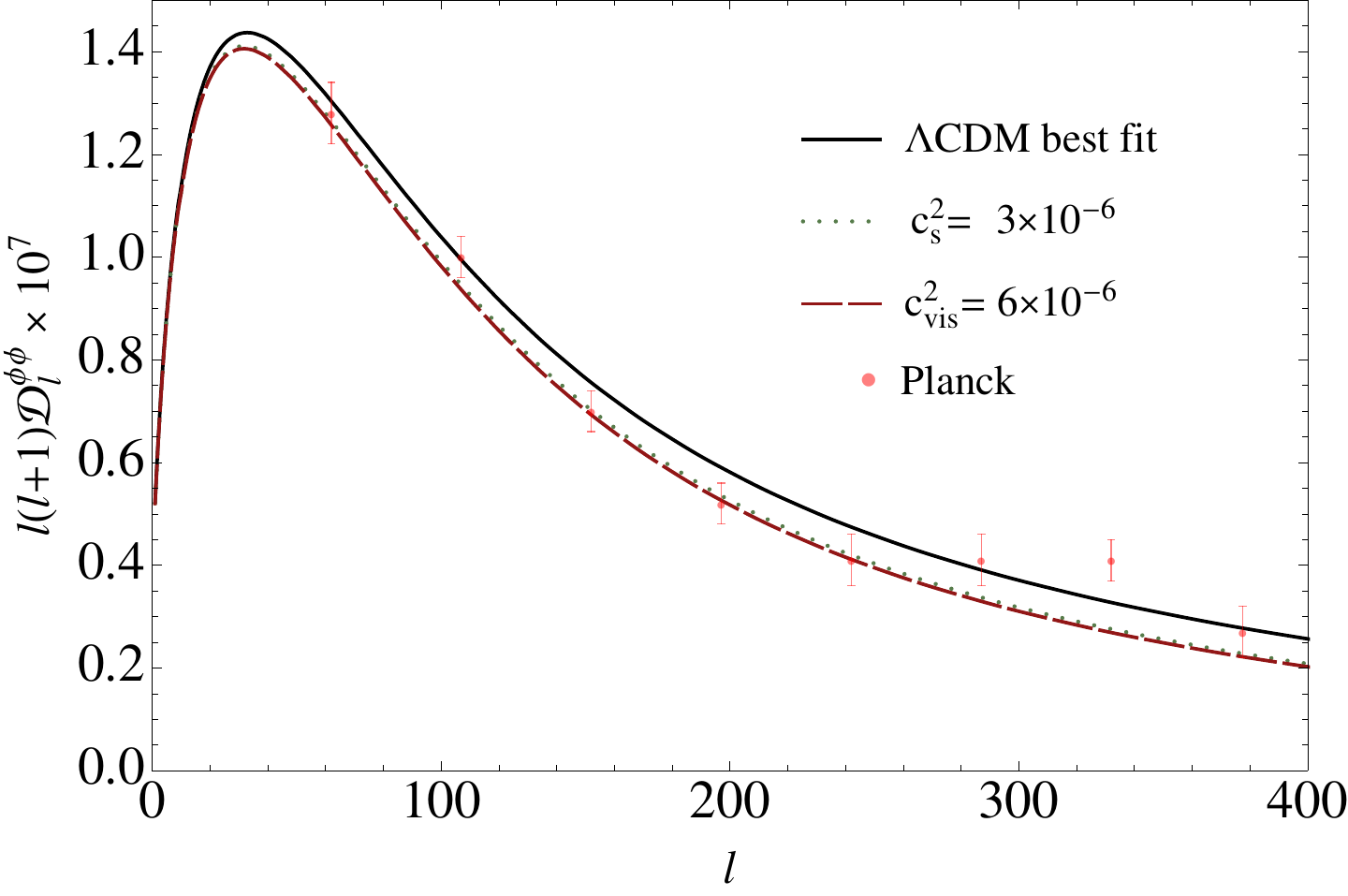}
 \caption{The lensing potential power spectrum plotted for the $\Lambda$CDM best-fit values and for values of $c^2_s$ and $c^2_\text{vis}$ that are close to our $99.7\%$ CL. The red points show the Planck data with error bars.
 }
\label{fig_cscv_PPchanges}
\end{figure}
\vspace{1mm}

The scale $k_\text{d}(\eta)$ at conformal time $\eta$ at which potential decay happens in a pure GDM universe is approximately
\begin{equation}
k^{-1}_\text{d}(\eta)=\eta \sqrt{c^2_s+\frac{8}{15}\cv^2}  \text{.}
\label{eq_k_dec}
\end{equation}
For length scales larger than $k_\text{d}^{-1}$, the effects of these two parameters on the gravitational potential are indistinguishable. For the scales relevant to the CMB, this induces 
a degeneracy between the two parameters.  Naively, the expression for $k_\text{d}$ suggests a negative correlation between the parameters such that the errors on $c^2_s$ 
should be about half the size of those on $c^2_\text{vis}$. This is approximately what is found for the errors in table \ref{table_w_speeds}. In addition, 
see Fig.~\ref{fig_2d_cv2_cs2}  where we show the 2-D contours in the $c^2_s$-$c^2_\text{vis}$ plane. In this figure, we have plotted lines that correspond 
to constant $k^{-1}_\text{d}$. The direction of these lines is a good fit to the direction of the contours, providing further evidence that this is the cause of the 
degeneracy between these parameters.

Since $c^2_s$ and $\cv^2$ do not affect the expansion history, the inclusion of either HST or BAO data has little effect on their constraints and this is precisely what is observed in
table \ref{table_w_speeds}.
However, the inclusion of the CMB lensing potential data (Lens) does have a significant effect on the $c^2_s$ and $\cv^2$ constraints, as opposed to what was found for the constraints on $w$. 
In particular, since potential decay in $\Lambda$-GDM after recombination leads to shallower lensing potentials, the result is a smaller lensing potential auto-correlation power spectrum plus a larger temperature-lensing potential cross-correlation due to the larger ISW contribution to the temperature spectrum.
As the lensing potential probes low redshifts, and the effects of the GDM parameters are cumulative over time, the effect on the lensing potential is stark.  
This is why the constraints on these parameters greatly improve when the Planck lensing data is included.

In figures \ref{fig_cscv_TTchanges} and \ref{fig_cscv_PPchanges}, we show the TT and lensing potential $C_l$s for 
the best-fit $\Lambda$CDM values, and also for values of $c^2_s$ and $c^2_\text{vis}$ that are close to our $99.7\%$ CL. In the TT plot, the residuals (differences between the GDM models and the best-fit $\Lambda$CDM model) have been multiplied by 50 in order to make them visible
as in Fig. \ref{fig_w_clchanges}.
 The lensing potential plot shows the significant change in the lensing potential due to GDM parameters. In the TT plot, the differences to $\Lambda$CDM also arise from lensing: the smaller lensing potential results in reduced smoothing of the peaks. These plots confirm that, for constant values of these parameters, it is predominantly the lensing that is generating the constraints. This could be different if the parameters were not constant. For example, the effect of the lensing would be smaller if the parameters scaled as $a^{-2}$ as is the case for warm dark matter \citep{LesgourguesTram2011}.  The values of $c^2_s$ and $c^2_\text{vis}$ here were chosen to reflect the $k_d$ degeneracy, and indeed there is little difference between the two GDM curves.

We can translate the upper bound on $c^2_s$ and $\cv^2$ into an upper bound on the ratio of $k_\text{d}^{-1}$ to the Hubble scale. 
At the $99.7\%$ CL, this is approximately $2.13\times10^{-3}$ (using $\eta\approx\adotoa^{-1}$), so the largest currently allowed scale on which GDM can modify cosmology is significantly below the Hubble scale. 
In this sense we consider our constraints on the GDM parameters to be strong.

\subsection{Constraints on the standard $\mathit{\Lambda}\!\,$CDM parameters.}
As expected, the inclusion of the GDM parameters worsens the constraints on some of the $\Lambda$CDM parameters, notably $\omega_g$, $\tau$ and the
derived parameters $H_0$ and $\sigma_8$, as seen in table~\ref{table_common_parameters}.

\begin{figure}
\centering
\begin{minipage}{3.5in}
  \centering
  \hspace{-0.3in}
\includegraphics[width=3.6in]{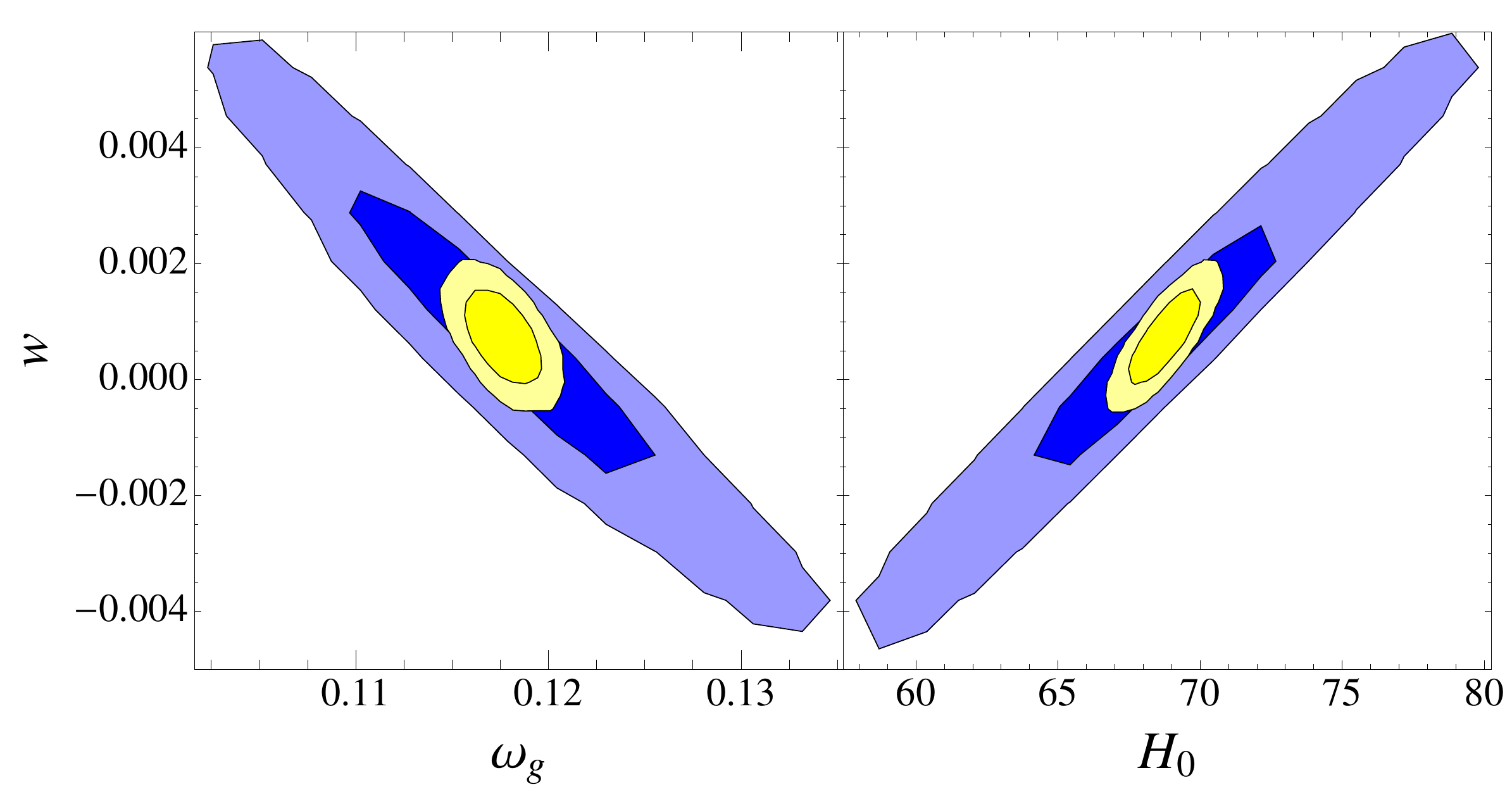}
 \caption{The $68.3\%$ and $95.5\%$ credible regions in the $w$-$\omega_g$ plane (left panel) and in the $w$-$H_0$ plane (right panel). 
 The blue (outer) contours are for the $\Lambda$-GDM run for PPS+Lens dataset combination, and the yellow (inner) contours are for the PPS+Lens+BAO dataset combination.
 The left panel shows a strong negative correlation between these parameters due to their effects 
 on the radiation-matter equality, whereas the right panel shows a strong positive correlation between these parameters 
due to their effect on the background expansion history. }
\label{fig_2d_w_omegag}
\end{minipage}
\end{figure}

The increased error bars for $\omega_g$ and $H_0$ are due to the strong degeneracies with $w$, as all three parameters
primarily affect the background expansion. See Fig.~\ref{fig_2d_w_omegag} for these degeneracies, which show the 2-D contours in the $w$-$\omega_g$ and $w$-$H_0$ 
 planes respectively. As explained in the previous subsection, $\omega_g$ and $w$ shift the radiation-matter equality which 
 in turn affects the heights of the first and second peaks in the CMB temperature spectrum (see \cite{KoppSkordisThomas2015}).
This results in a negative correlation between the two parameters, hence  the degeneracy seen on the left panel of Fig.~\ref{fig_2d_w_omegag}.

As changing $w$ also has an effect on the expansion history, we expect to get a degeneracy with $H_0$. In particular, 
as is well known $H_0$ and $\omega_g$ are negatively correlated even in $\Lambda$CDM \citep{HinshawEtAl2012}
and since  $w$ and $\omega_g$ are also negatively correlated, we expect $w$ and $H_0$  to be positively correlated. 
Indeed this is verified on the right panel of  Fig.~\ref{fig_2d_w_omegag}.
\begin{figure}
  \centering
   \includegraphics[width=3.4in]{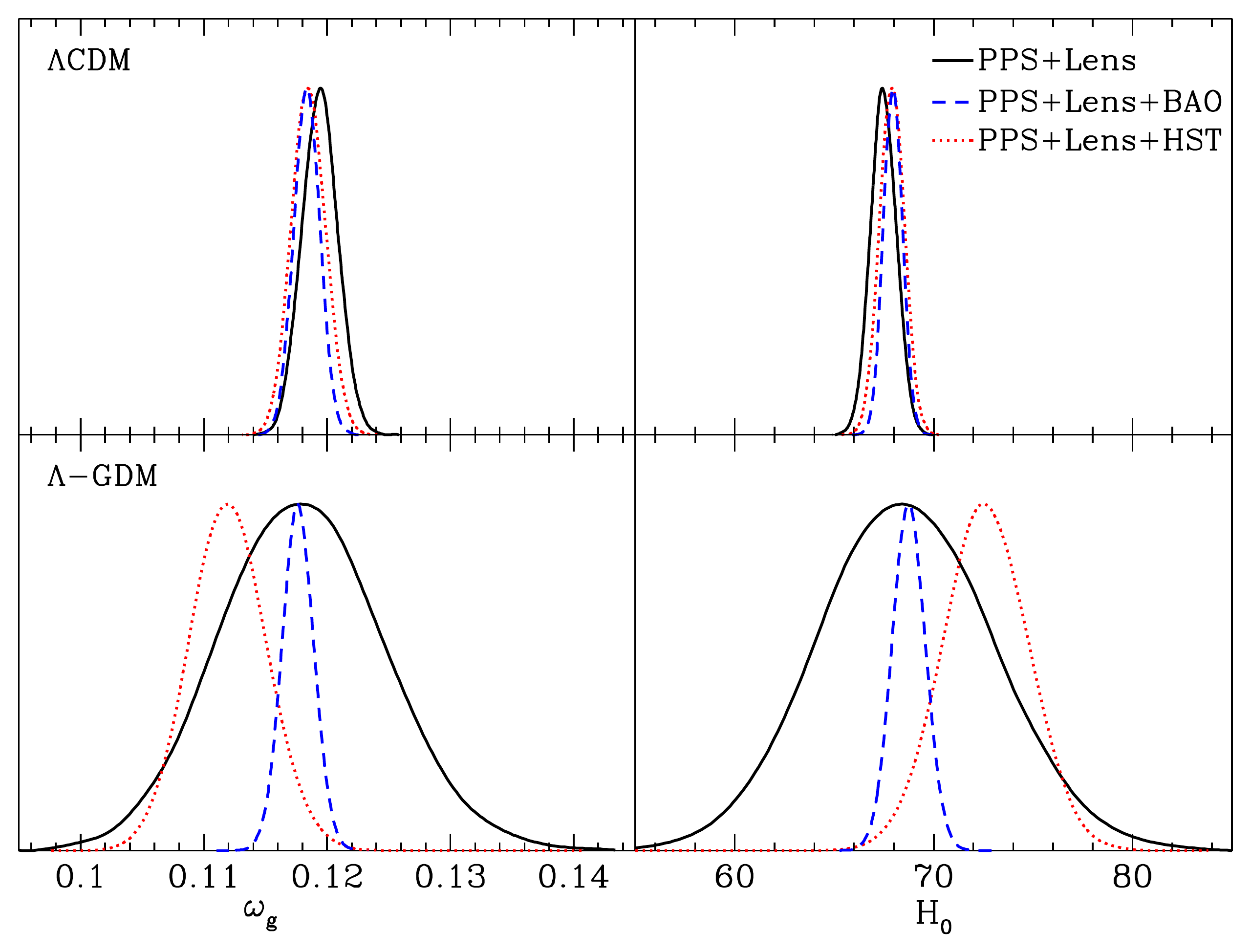}
\caption{The 1d posteriors for $\omega_g$ (left panel) and $H_0$ (right panel) in the case of $\Lambda$CDM. The black (solid) curve is with the combination of  PPS + Lens,
 the blue (dashed) curve is with PPS + Lens + BAO and the red (dotted) curve is for PPS + Lens with the addition HST prior.}
\label{fig_4plots}
\end{figure}

The full 1-D posteriors for the PPS + Lens dataset combination for $\omega_g$ and $H_0$ can be found in Fig.~\ref{fig_4plots}. 
In this plot, the black (solid) curves show the posteriors for the $\Lambda$CDM model (upper panel) and $\Lambda$-GDM model (lower panel) respectively.
 In addition to the broadening of the constraints, there is also a slight shift 
away from the mean value found in the $\Lambda$CDM model, however, this shift remains within the $68.3\%$ credible region for both parameters. The posterior
 for $\Lambda$-wDM is similar to the  $\Lambda$-GDM case~\footnote{albeit with a slightly greater shift in the mean} as is expected from the lack 
of correlation of the perturbative GDM parameters with either $w$ or $\omega_g$ as shown in  Fig.~\ref{fig_2d_w_cs2},
 and we choose not to plot it.

As the $\omega_g$ and $H_0$ parameters primarily affect the background expansion history, their constraints are influenced by the inclusion of either the HST or BAO data. 
In Fig.~\ref{fig_4plots} we also plot the 1-D posteriors for $\omega_g$ and $H_0$, for the combination of PPS + Lens with the inclusion of the HST prior, depicted by the red (dotted) curve.
Once again, there is little difference for these parameters between the $\Lambda$-wDM and $\Lambda$-GDM models and we do not plot the former.
We see that the addition of the HST prior significantly improves the constraints on these parameters. There is still an offset of the mean value of the posterior 
compared to $\Lambda$CDM, which is now more significant due to the reduced width of the posteriors.

Fig.~\ref{fig_4plots} also shows the posteriors for the inclusion of the BAO data rather than the HST prior, depicted by the blue (dashed) curve. 
For this combination, the constraints on the parameters are even tighter, with the constraints from the $\Lambda$-GDM model (and similarly for the $\Lambda$-wDM model which is not plotted) 
 now being almost as strong as those from the $\Lambda$CDM runs, despite the extra parameters. Moreover, for this combination of datasets, 
there is no significant offset of the means compared to $\Lambda$CDM. The shrinking of the $w$-$\omega_g$ and $w$-$H_0$ contours with the inclusion of BAO
 is also clearly visible in Fig. \ref{fig_2d_w_omegag}.

\begin{figure}
  \centering
 \includegraphics[width=3.4in]{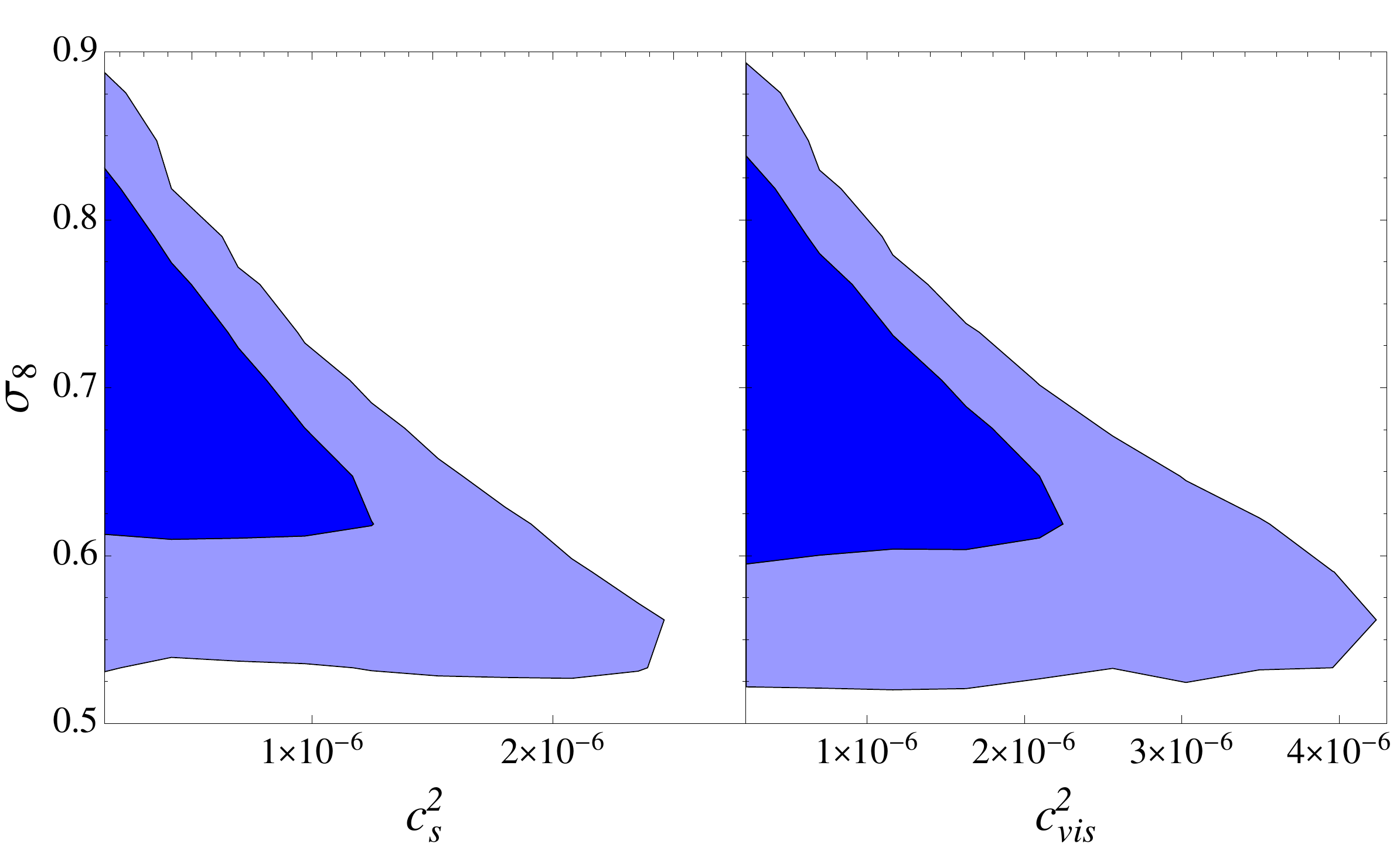}
 \caption{The $68.3\%$ and $95.5\%$ credible regions of $\sigma_8$ with the perturbative GDM parameters $c_s^2$ (left) and $\cv^2$ (right) for the  $\Lambda$-GDM model using the PPS + Lens dataset combination. 
The strong negative correlation in these two pairs of parameters is due to the strong reduction in structure growth induced by the GDM parameters.
}
\label{fig_2d_sig8_cs2}
\end{figure}

In the $\Lambda$-wDM model, $w$ has a degeneracy with $\sigma_8$. However, in the $\Lambda$-GDM model, this is subdominant to the stronger degeneracy between $\sigma_8$ and the perturbative parameters $c^2_s$ and $\cv^2$. 
This is due to the strong effect that these two parameters have on the growth of structure, as found in \cite{KoppSkordisThomas2015}. 
These two parameters both greatly reduce the growth of the matter perturbations on length scales below $k_\text{d}^{-1}$. Since $\sigma_8$ is defined to be the amplitude of this 
spectrum, this results in a strong degeneracy between these two GDM parameters and $\sigma_8$, which is sufficiently strong to replace the usual degeneracy between $\sigma_8$ and $A_s$. 
In Fig.~\ref{fig_2d_sig8_cs2}, we show the 2D contours in the $\sigma_8$-$c^2_s$ and $\sigma_8$-$\cv^2$ planes respectively. Fixing either of $\sigma_8$ and $c^2_s$ (or $\cv^2$), 
and increasing the other, results in $A_s$ increasing. Thus we see a negative correlation between $\sigma_8$ and these GDM parameters.

\begin{figure}
  \centering
   \includegraphics[width=3.4in]{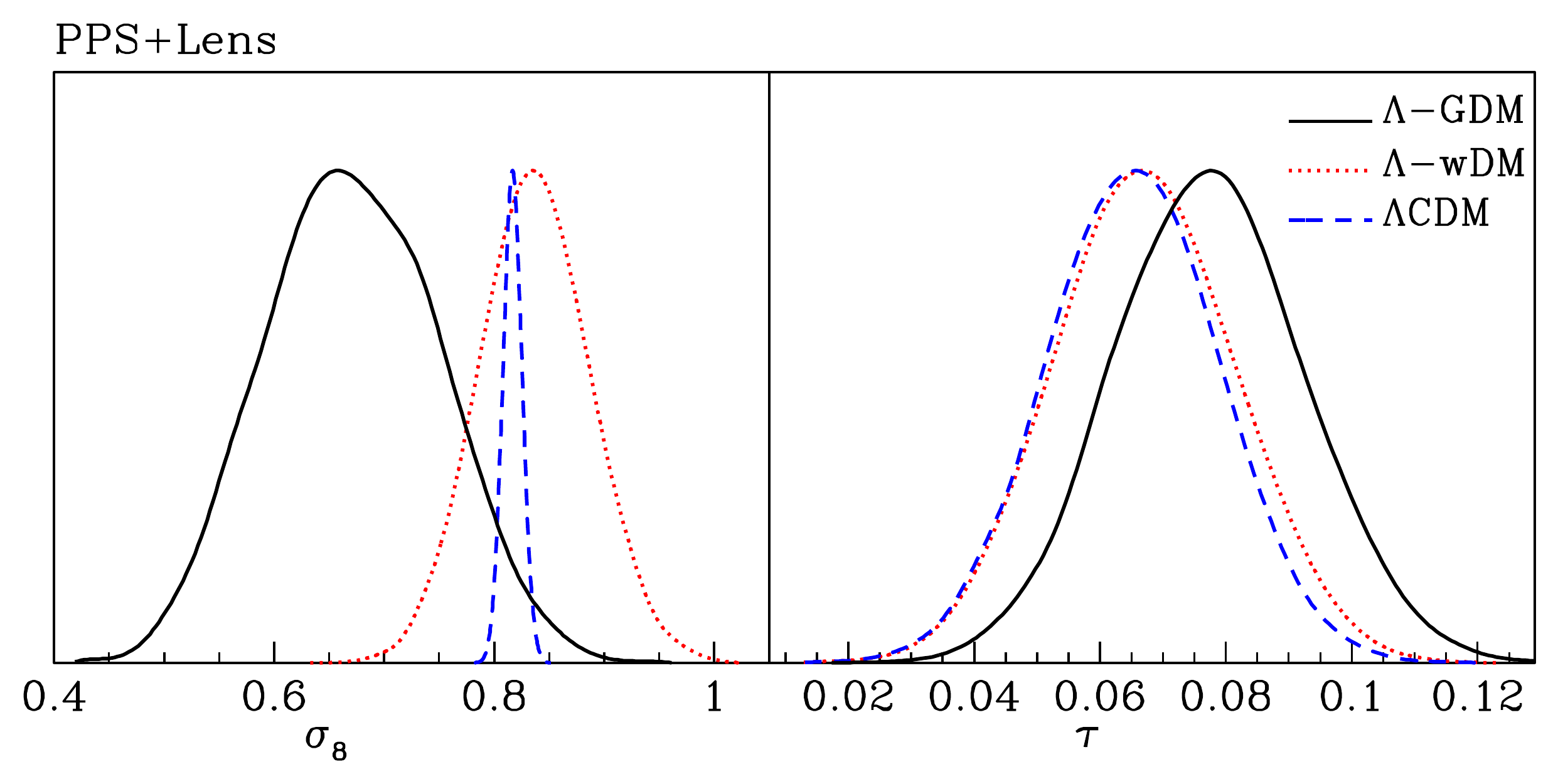}
\caption{The 1d posteriors for $\sigma_8$ (left panel) and $\tau$ (right panel) for the combination of Planck Primary and Planck Lensing data; the blue (dashed) curve is for $\Lambda$CDM,
the red (dotted) curve is for the $\Lambda$-wDM run and the black (solid) curve is for the $\Lambda$-GDM run. }
\label{fig_2plots_sig8tau}
\end{figure}

It is interesting to look more closely at $\sigma_8$ for the the combination of Planck + Planck lensing data. The 1-dimensional posteriors are shown in Fig.~\ref{fig_2plots_sig8tau} 
for $\Lambda$CDM (blue, dashed), $\Lambda$-wDM (red, dotted) and $\Lambda$-GDM (back, solid). From this plot it is easily seen that, although the errors increase
 significantly for the $\Lambda$-wDM run, the mean value stays close to the $\Lambda$CDM value. In contrast, in the $\Lambda$-GDM run, in addition to the 
increase in the errors, the mean shifts compared to $\Lambda$CDM. We saw before that increasing the GDM parameters reduces the value of $\sigma_8$ for fixed cosmological parameters 
(including fixed $A_s$). Since these two GDM parameters only take positive values,  $\sigma_8$ can only be reduced, i.e. it becomes biased towards smaller numbers,
 relative to the $\Lambda$CDM value. Thus, the inclusion
 of these two parameters results in the posteriors being shifted, even though the GDM parameters themselves are found to be consistent with zero.

The mean value for the parameter $\tau$ is affected by the inclusion of the GDM parameters for the combination of PPS + Lens data.
The 1-dimensional posteriors for $\tau$ from the different models are also plotted in Fig.~\ref{fig_2plots_sig8tau}. We can see that for the $\Lambda$-wDM model,
 the mean and width of the posterior have changed little from the $\Lambda$CDM values.  For the $\Lambda$-GDM case, the width of the posterior is 
very similar to the $\Lambda$CDM value, however, the mean value has increased, albeit by less than the $68.3\%$ CL constraint.
Since $\tau$ and $A_s$ are positively correlated, while $A_s$ and $\sigma_8$ are anti-correlated, 
we expect $\tau$ and $\sigma_8$ to be anti-correlated. Therefore $\tau$ is biased for the same reason as $\sigma_8$ but in the opposite direction.

Before closing this discussion let us remark that we also ran chains with the low-$l$ Planck likelihood and the TT only high-$l$ likelihood, 
again with the complete ``not-lite'' set of nuisance parameters. 
For these runs the constraints on $c^2_s$ and $c^2_\text{vis}$ become about $50\%$ worse compared to the PPS run,
as expected from the reduction in information. The constraints on $w$ worsen significantly, as do those of the parameters that it
 has its key degeneracies with ($\omega_g$ and $H_0$, as discussed above). This is because there is simply not enough information in the temperature spectrum 
to constrain the expansion history once the extra degeneracies introduced by $w$ are included. The extra information provided by the polarisation 
is sufficient to constrain the $\Lambda$CDM parameters to values that are approximately in line with Planck, with the exceptions noted above.

\subsection{Comparison to previous work}
\label{sec_comparison}
Although a full comparison of the GDM parameters as a replacement for CDM has not been previously performed, several
works have looked at including an equation of state for dark matter.

One of the first works to constrain the dark matter equation of state with cosmological data was \cite{Muller2005}. In that work, two cases for $w$ were
constrained using a combination of background data and matter power spectrum data. Their first case corresponds to setting $w=c^2_s$ in our notation 
(note that this means a negative value of $c^2_s$ was allowed in their analysis, which is unphysical).
For that case, they found strong  constraints on $w$, $\sim 10^{-6}$ at the $99.7\%$ CL, due to the strong effect of the sound speed on matter clustering 
and the inclusion of the matter power spectrum data to constrain $\sigma_8$. These constraints are comparable to those obtained in this work on $c^2_s$, as would be expected. 
Their second case is similar to our $\Lambda$-wDM case, although not identical due to the difference in the definition
of the non-adiabatic pressure. In that case the constraints are much more non-Gaussian than ours, with $-8.78\times10^{-3}<w<1.86\times10^{-3}$ at the
$99.7\%$ CL. Their constraints are at a similar level to those obtained in this paper with PPS+Lens+HST data, and less constraining than
 the tightest constraints in this paper obtained with the PPS+Lens+BAO dataset combination.

The dark matter equation of state (the $\Lambda$-wDM model) was constrained in \cite{CalabreseMigliaccioMelchiorriEtal2009} using WMAP\citep{HinshawWeilandHilletal2009} data. 
They found $w$ to be consistent with zero, with the constraints from WMAP alone being  $(-0.35^{+1.17}_{-0.98})\times 10^{-2}$ at the $95.5\%$ CL, which is more than 
a factor of two worse than those we obtain using PPS or  PPS + Lens. The inclusion of additional datasets including extra CMB datasets, SNLS supernovae 
data \citep{AstierGuyRegnaultEtAl2006} and SDSS matter power spectrum data \citep{TegmarkEisensteinStraussEtAl2006} improved their constraints to $(0.07^{+0.41}_{-0.42})\times10^{-2}$ 
at the $95.5\%$ CL, which is very similar to the constraints we obtain from PPS + Lens. The strongest constraints in this paper, obtained using PPS + Lens + BAO data, 
are approximately 3 times tighter than the tightest constraints in \cite{CalabreseMigliaccioMelchiorriEtal2009}.

One of the more well-known models that fits under the GDM framework is the Chaplygin gas, and this model has been compared to Planck data previously
\citep{LiXu2014}. However, due to the difference between the forms of the GDM parameters in the Chaplygin gas model and those adopted in this work, our constraints are not
directly comparable.

Two works \citep{WeiChenLiu2013,KumarXu2012} examine the equation of state of dark matter using a Taylor expansion around
$a=1$ as $w_m =w_{m0} +w_{ma}(1-a)$.
In \cite{WeiChenLiu2013}, the authors constrain the equation of state of dark matter using only background quantities, obtaining error bars that are typically larger than those obtained here.
In \cite{KumarXu2012}, further data sets including Planck and matter power spectrum measurements are also included, and the resulting error bars are typically slightly smaller than those obtained here, albeit highly non-Gaussian.
However, as the equation of state was parameterised differently to our work, the results are not easy to compare directly to ours.

In a series of papers \citep{Xu2014,XuChang2013}, the authors constrain the equation of state of dark matter (the $\Lambda$-wDM model) using different datasets 
in addition to Planck, including BAO data and SNLS3 \citep{GuySullivanConleyEtAl2010,SullivanGuyConleyEtAl2011} data to constrain the background expansion, 
as well as examining the effect of including WIGGLEz matter power spectrum data \citep{ParkinsonRiemer-SorensonBlakeEtAl2012}. At the $99.7\%$ CL, the 
constraints are found to be $\sim 2\times10^{-3}$, which is very similar to the constraints on $w$ that we find here for the combination PPS + Lens + BAO. 
The inclusion of the WIGGLEz data has only a small effect on the constraints on $w$, again similarly to what was found in this paper with the effect on $w$ of 
including the Planck Lensing (Lens) likelihood.

\section{Conclusion and discussion}
\label{sec_conclusion}
The main results of this paper comprise tables \ref{table_w_speeds} and  \ref{table_common_parameters}. These show the results of an MCMC analysis of the Planck satellite data,
where the standard $\Lambda$CDM model has been extended to include the three (constant) GDM parameters, namely the equation of state $w$, the sound speed $c_s^2$ 
and viscosity $\cv^2$ of the dark matter fluid. We found that these additional parameters are all consistent with zero. and are strongly constrained
from the combination of high-$l$ and low-$l$ Planck likelihoods. The inclusion of Planck lensing data, and either BAO data or the HST prior, further tightens the constraints.
The equation of state $w$ is constrained to be  $-0.000896<w<0.00238$ at the $99.7\%$ CL and the parameters $c^2_s$ and $\cv^2$ are constrained to be less 
than $3.21\times10^{-6}$ and $6.06\times10^{-6}$ respectively at the $99.7\%$ CL. We found that the CMB lensing likelihood is important for constraining 
the parameters $c^2_s$ and $\cv^2$, whereas the inclusion of an additional dataset that constrains the background expansion of the universe, either BAO or HST, is important for constraining $w$.

We uncovered a degeneracy between the two perturbative GDM parameters $c^2_s$ and $\cv^2$ which remains for all types of datasets used in this work. This is due to
 the way these two parameters affect the decay of the gravitational potential on length scales smaller than $k_{\text{d}}^{-1}$ as in (\ref{eq_k_dec}).  However, 
 this degeneracy is broken well below the scale $k^{-1}_\text{d}$. In this regime, for a fixed $k_\text{d}$, $c^2_s$ causes a faster decay of the potential as well as resulting in
 oscillations 
of the matter power spectrum that are not present in a universe with $c^2_s=0$ and $\cv^2\neq0$. 
This suggests that including additional data on the matter power spectrum at late times, typically at even lower redshift than that probed by the 
CMB lensing potential, could improve the constraints on these parameters further. In addition, data on the matter power spectrum may probe the region 
where $c^2_s$ and $\cv^2$ have different effects, thus breaking degeneracies. We intend to investigate this in future work.

We have also examined the effects of including the GDM parameters on the constraints of the standard $\Lambda$CDM parameters. For the full Planck dataset 
including lensing potential reconstruction, the main effect is to significantly increase the error bars on $\omega_g$ as well as on the derived parameters $\sigma_8$ and $H_0$.
 The parameters $c^2_s$ and $\cv^2$ inhibit the growth of structure below the length scale $k_\text{d}^{-1}$, reducing the matter power spectrum on those scales. 
Thus, the inclusion of $c^2_s$ and $\cv^2$ loosens the constraint on $\sigma_8$, as well as shifting the posterior to smaller values. This shift is due to the requirement 
that these parameters be non-negative, therefore they can only act to decrease $\sigma_8$ (and not to increase it) relative to $\Lambda$CDM. 
The parameter $\tau$ 
is affected as well. When only $w$ is added to the standard $\Lambda$CDM parameters, the posterior remains similar to that for the $\Lambda$CDM analysis. 
However, when all three GDM parameters are included, the mean of the 
posterior is shifted to higher values. We note that the constraints on $\tau$ are nonetheless similar for all three cases.

We examined the effect of using HST and BAO data in addition to the Planck dataset. These improved the constraints on $w$, with the BAO data having a larger
effect than the HST prior. For both the $\Lambda$-wDM and $\Lambda$-GDM models these additional datasets significantly improved the constraints on $\omega_g$ and $H_0$. Again, the inclusion of the BAO data has a larger effect, such that in this case $H_0$ and $\omega_g$ are nearly as well constrained as in a $\Lambda$CDM cosmology without the additional GDM parameters.

We have only considered the case of constant values of the GDM parameters in this work. This should act as a null test for whether there are any significant effects on the 
CMB from dark matter properties. Nonetheless, it is possible that a more sophisticated parameterisation, perhaps following a specific model, may result in 
non-zero values of the GDM parameters being preferred. In addition, we note that in our analysis we have not included
many of the additional parameters that are unnecessary in a $\Lambda$CDM analysis, such as the curvature $\Omega_k$ and isocurvature perturbations.
It may be that the inclusion of these parameters would  allow for a non-zero value of the GDM parameters, which, combined with the effects on $\sigma_8$,
may allow the tension between Planck and observations of the late universe to be resolved. 
We leave this to future work.

\acknowledgements
We thank C. B{\oe}hm for useful discussions and T. Tram and B. Audren for help regarding the CLASS and MontePython codes. We also thank all of  the authors of these codes for making them publicly available.
The research leading to these results 
has received funding from the European Research Council under the European Union's Seventh Framework Programme (FP7/2007-2013) / ERC Grant Agreement n. 617656 ``Theories
 and Models of the Dark Sector: Dark Matter, Dark Energy and Gravity''.

\bibliographystyle{unsrtnatetal}
\bibliography{MCMC_I.bib}

\setlength{\tabcolsep}{1mm}
\begin{table*}
 \caption{Constraints}
\centering
\label{table_common_parameters}
\begin{mytabular}[1.8]{|c|c||c|c|c|c|c|c|}
\hline
\hline
 &  Likelihoods  & \multicolumn{2}{|c|}{PPS+Lens}  & \multicolumn{2}{|c|}{PPS+Lens+HST}  & \multicolumn{2}{|c|}{PPS+Lens+BAO} \\
 \cline{2-8}
Models & Parameters &   $95.5\%$ CL & $99.7\%$ CL &  $95.5\%$ CL & $99.7\%$ CL &  $95.5\%$  CL & $99.7\%$ CL  \\
\hline 
               & $\omega_b$         & $0.02225^{+0.00032}_{-0.00032}$ & $0.02225^{+0.00049}_{-0.00048}$ & $0.02233^{+0.00032}_{-0.00031}$ & $0.02233^{+0.00047}_{-0.00045}$ & $0.02233^{+0.00029}_{-0.00029}$ & $0.02233^{+0.00042}_{-0.00043}$ \\
               & $\omega_g$            & $0.1194^{+0.0030}_{-0.0029}$ & $0.1194^{+0.0045}_{-0.0042}$ & $0.1185^{+0.0028}_{-0.0028}$ & $0.1185^{+0.0042}_{-0.0041}$ & $0.1184^{+0.0022}_{-0.0022} $& $0.1184^{+0.0032}_{-0.0032}$  \\
               & $100\theta_s$         & $1.04184^{+0.00061}_{-0.00061}$ & $1.04184^{+0.00089}_{-0.00090}$ & $1.04193^{+0.00060}_{-0.00060}$ & $1.04193^{+0.00089}_{-0.00087}$ & $1.04194^{+0.00059}_{-0.00059}$ & $1.04194^{+0.00087}_{-0.00088}$ \\
               & $\ln 10^{10} A_s$     & $3.064^{+0.051}_{-0.051}$ & $3.064^{+0.076}_{-0.076}$ & $3.074^{+0.051}_{-0.052}$ & $3.074^{+0.075}_{-0.077}$ & $3.075^{+0.046}_{-0.047}$ & $3.075^{+0.068}_{-0.069}$ \\
 $\Lambda$CDM  & $n_s$                 & $0.9646^{+0.0099}_{-0.0097}$ & $0.9646^{+0.015}_{-0.014}$ & $0.9670^{+0.0097}_{-0.0096}$ & $0.9670^{+0.014}_{-0.014}$ & $0.9673^{+0.0084}_{-0.0084}$ & $0.9673^{+0.013}_{-0.012}$ \\
               & $\tau$                & $0.065^{+0.028}_{-0.028}$ & $0.065^{+0.042}_{-0.042}$ & $0.0713^{+0.028}_{-0.028}$ & $0.0713^{+0.041}_{-0.042}$ & $0.072^{+0.025}_{-0.025}$ & $0.072^{+0.037}_{-0.037}$ \\
               & $\Omega_\Lambda $     & $0.687^{+0.018}_{-0.018}$ & $0.687^{+0.026}_{-0.028}$ & $0.693^{+0.017}_{-0.017}$ & $0.693^{+0.024}_{-0.026}$ & $0.694^{+0.013}_{-0.013}$ & $0.694^{+0.019}_{-0.019}$ \\
               & $H_0$                 & $67.5^{+1.3}_{-1.3}$ & $67.5^{+2.0}_{-2.0}$ & $67.9^{+1.3}_{-1.3}$ & $67.9^{+1.9}_{-1.9}$ & $70.0^{+1.0}_{-1.0}$ & $68.0^{+1.5}_{-1.5}$ \\
               & $\sigma_8 $           & $0.817^{+0.018}_{-0.018}$ & $0.817^{+0.027}_{-0.026}$ & $0.818^{+0.018}_{-0.018}$ & $0.818^{+0.027}_{-0.027}$ & $0.819^{+0.017}_{-0.018}$ & $0.819^{+0.026}_{-0.027}$ \\
\hline
                & $\omega_b$         
& $0.02223^{+0.00034}_{-0.00033}$ & $0.02223^{+0.00051}_{-0.00049}$ &
                 $0.02221^{+0.00033}_{-0.00033}$ & $0.02221^{+0.00049}_{-0.00048}$ & $0.02224^{+0.00033}_{-0.00033}$ & $0.02224^{+0.00047}_{-0.00049}$  \\
                & $\omega_g$            
& $0.1170^{+0.0138}_{-0.0134}$ & $0.1170^{+0.0212}_{-0.0189}$ &
                 $0.1117^{+0.0064}_{-0.0062}$ & $0.1117^{+0.0095}_{-0.0092}$ & $0.1175^{+0.0027}_{-0.0026}$ &  $0.1175^{+0.0040}_{-0.0039}$ \\
                & $100\theta_s$         
 & $1.04189^{+0.00065}_{-0.00067}$ & $1.04189^{+0.00099}_{-0.00099}$ &
                 $1.04201^{+0.00062}_{-0.00061}$ & $1.04201^{+0.00093}_{-0.00091}$ & $1.04188^{+0.00060}_{-0.00060}$ & $1.04188^{+0.00088}_{-0.00089}$  \\
                & $\ln 10^{10} A_s$  
& $3.068^{+0.057}_{-0.056}$ & $3.068^{+0.082}_{-0.083}$ &
                 $3.080^{+0.051}_{-0.051}$ & $3.080^{+0.075}_{-0.077}$ & $3.067^{+0.048}_{-0.049}$ & $3.067^{+0.072}_{-0.075}$  \\
  $\Lambda$-wDM & $n_s$                
& $0.9665^{+0.0139}_{-0.0140}$ & $0.9665^{+0.0207}_{-0.0204}$ &
   $0.9709^{+0.0102}_{-0.0101}$ & $0.9709^{+0.0153}_{-0.0146}$ & $0.9658^{+0.0087}_{-0.0086}$ & $0.9658^{+0.0127}_{-0.0131}$  \\
                & $\tau$                
& $0.067^{+0.030}_{-0.029}$ & $0.067^{+0.044}_{-0.043}$ &
                 $0.072^{+0.028}_{-0.027}$ & $0.072^{+0.041}_{-0.041}$ & $0.067^{+0.027}_{-0.027}$ & $0.067^{+0.040}_{-0.040}$  \\
                & $\Omega_\Lambda $     
& $0.703^{+0.102}_{-0.112}$ & $0.703^{+0.128}_{-0.187}$ &
                 $0.745^{+0.042}_{-0.043}$ & $0.745^{+0.058}_{-0.069}$ & $0.703^{+0.020}_{-0.020}$ & $0.703^{+0.029}_{-0.031}$  \\
                & $H_0$                 
& $69.3^{+9.0}_{-9.1}$ & $69.3^{+13.4}_{-13.4}$ & $72.8^{+4.3}_{-4.4}$ & $72.8^{+6.5}_{-6.4}$ & $68.8^{+1.7}_{-1.7}$ & $68.8^{+2.5}_{-2.5}$  \\
                & $\sigma_8 $          
& $0.837^{+0.103}_{-0.102}$ & $0.837^{+0.155}_{-0.149}$ & $0.876^{+0.053}_{-0.052}$ & $0.876^{+0.079}_{-0.077}$ & $0.831^{+0.028}_{-0.028}$ & $0.831^{+0.043}_{-0.041}$  \\
\hline
                & $\omega_b$        
& $0.02219^{+0.00033}_{-0.00033}$ & $0.02219^{+0.00049}_{-0.00049}$ &
                 $0.02216^{+0.00034}_{-0.00033}$ & $0.02216^{+0.00052}_{-0.00049}$ & $0.02218^{+0.00033}_{-0.00033}$ & $0.02218^{+0.00048}_{-0.00049}$  \\
                & $\omega_g$            
& $0.1180^{+0.0135}_{-0.0134}$ & $0.1180^{+0.0205}_{-0.0194}$ &
                 $0.1120^{+0.0062}_{-0.0062}$ & $0.1120^{+0.0095}_{-0.0092}$ & $0.1176^{+0.0027}_{-0.0027}$ &  $0.1176^{+0.0041}_{-0.0040}$ \\
                & $100\theta_s$  
& $1.04186^{+0.00065}_{-0.00064}$ & $1.04186^{+0.00096}_{-0.00097}$ &
                 $1.04199^{+0.00060}_{-0.00060}$ & $1.04199^{+0.00089}_{-0.00088}$ & $1.04186^{+0.00059}_{-0.00059}$ & $1.04186^{+0.00090}_{-0.00091}$  \\
                & $\ln 10^{10} A_s$  
& $3.090^{+0.060}_{-0.059}$ & $3.090^{+0.090}_{-0.087}$ &
                 $3.100^{+0.057}_{-0.056}$ & $3.100^{+0.084}_{-0.083}$ & $3.089^{+0.057}_{-0.055}$ & $3.089^{+0.084}_{-0.080}$  \\
  $\Lambda$-GDM  & $n_s$                 
& $0.9651^{+0.0140}_{-0.0136}$ & $0.9651^{+0.0203}_{-0.0205}$ &
                 $0.9698^{+0.0103}_{-0.0104}$ & $0.9698^{+0.0156}_{-0.0151}$ & $0.9651^{+0.0087}_{-0.0087}$ & $0.9651^{+0.0131}_{-0.0123}$  \\
                & $\tau$                
& $0.0773^{+0.0310}_{-0.0303}$ & $0.0773^{+0.0456}_{-0.0445}$ &
                 $0.0815^{+0.0298}_{-0.0295}$ & $0.0815^{+0.0450}_{-0.0440}$ & $0.0769^{+0.0297}_{-0.0293}$ & $0.0769^{+0.0441}_{-0.0433}$  \\
                & $\Omega_\Lambda $     
& $0.695^{+0.102}_{-0.112}$ & $0.695^{+0.134}_{-0.197}$ &
                 $0.743^{+0.041}_{-0.044}$ & $0.743^{+0.059}_{-0.070}$ & $0.703^{+0.020}_{-0.020}$ & $0.703^{+0.029}_{-0.031}$  \\
                & $H_0$                 
& $68.6^{+8.9}_{-8.8}$ & $68.6^{+13.6}_{-12.7}$ &
                 $72.6^{+4.3}_{-4.3}$ & $72.6^{+6.4}_{-6.5}$ & $68.8^{+1.7}_{-1.7}$ & $68.8^{+2.6}_{-2.6}$  \\
                & $\sigma_8 $          
 & $0.671^{+0.155}_{-0.155}$ & $0.671^{+0.226}_{-0.213}$ &
                 $0.702^{+0.154}_{-0.156}$ & $0.702^{+0.197}_{-0.216}$ & $0.672^{+0.134}_{-0.143}$ & $0.672^{+0.167}_{-0.191}$  \\
\hline
\hline
\end{mytabular}
\end{table*}

\end{document}